\documentstyle[preprint,aps]{revtex}
\begin{document}
\draft
\title{
Eigenvalue Correlations for Banded Matrices}
\author{Pragya Shukla$^{*}$} 
\address{Department of Physics,   
Indian Institute of Technology, Kharagpur, India.}

\maketitle
\begin{abstract}
	We study the evolution of the distribution of eigenvalues of a 
$N\times N$ matrix ensemble subject to a change of variances of its 
 matrix elements.
 Our results indicate that the evolution of the probability density is    
 governed by a Fokker-Planck equation similar 
to the one governing the time-evolution of the particle-distribution in 
Wigner-Dyson gas, with relative variances now playing the role of time. 
This is also similar to the Fokker-Planck equation for the   
distribution of eigenvalues of a $N \times N$ matrix subject to a random 
perturbation taken from the standard Gaussian ensembles with perturbation 
strength as the "time" variable. This  equivalence alongwith the already   
known correlations of standard Gaussian ensembles can therefore help 
us to obtain the correlations for various physically-significant cases 
modeled by random banded Gaussian ensembles. 
\end{abstract}
.

	It is now well-established that the statistics of energy levels 
of complex quantum systems e.g. chaotic systems, disordered systems 
 can be well-modeled by the ensembles of 
 large random matrices \cite{gmw}. The nature of underlying quantum dynamics of these 
 systems  divides the applicable random matrix ensembles (RME) into two 
 major classes. The ensembles of the full random matrices (RFME) \cite{meta} have been very 
successful in modelling the statistical properties of systems with 
delocalized quantum dynamics. On the other hand, presence of the localized 
quantum dynamics requires a consideration of the Band random matrices 
ensembles (RBME) which can generally be described as a 
$N\times N$ matrix with non-zero elements effectively within a band of the 
width $b>1$ around the main diagonal. 
 The presence of these two major 
classes can be understood as follows. In the matrix representation of an 
operator, an off-diagonal matrix element describes the overlapping of the 
eigenfunctions in the basis space. For the delocalized quantum dynamics, the 
eigenfunctions are extended and the overlapping between them  
and therefore off-diagonal matrix elements are of the same order. The ensembles 
of such matrices are referred here as the RFME.  In contrast, the localization 
of the eigenfunctions implies a decaying tendency of overlapping between them,  
 leading to a diminishing strength of the  corresponding off-diagonal 
matrix elements and therefore RBM. Our aim in this paper is to suggest a 
general method for the statistical studies of RBME.

	Various studies of the eigenfunctions of complex systems have 
revealed the existence of various types of localizations, intermediate 
between fully localized eigenfunctions (occurs only when system size 
$L\rightarrow \infty$) and fully extended eigenstates (implying that the 
localization length $\ge$ the system size). For example, the eigenfunctions, 
in a weakly disordered potential exhibiting the Anderson metal-insulator 
transition, are essentially structureless and extended and the statistics 
here can be well-modeled by FRME. In the insulator regime, a complete 
localization of eigenfunctions leads to a zero correlation among them 
and a Poisson statistics for energy levels.  
	However	the eigenfunctions in the critical region near 
the Anderson transition reveal a special feature "multifractality" in 
their structure \cite{km}.
 The multifractal nature of a wavefunction is commonly accepted
to be related to its space structure which  
 also reveals itself in their
overlapping with each other; 
the two fractal wavefunctions, however
sparse they are, overlap strongly in contrast to two fully localized states. 
The nature of the energy level statistics being related to the 
structure of eigenfunctions as well as their overlapping,
the latter's special feature should manifest
also in the former. Indeed the presence of multifractality  
leads to a new type of universal critical level statistics which
is described by a set of critical exponents  
 and is different from both Wigner-Dyson  (FRME) as well as 
Poisson statistics \cite{sssls,kla}.  
 Recently there have
been suggestions regarding the existence of such  
eigenfunctions in wide variety of physical systems. For example, a
Coulomb impurity inside an integrable square billiard leads to multifractal  
 eigenfunctions in momentum representation 
however small is the strength of the potential \cite{al}. It also  
seems quite possible that the multifractality 
is a quantum mechanical manifestation of the special nature of
underlying classical dynamics, intermediate between integrability 
(fully localized eigenfunctions) and chaos 
(fully extended eigenfunctions) \cite{km}.

Recent research on complex systems has indicated that 
RBME, with zero mean value of all the matrix elements and 
 the variance $<|H_{ij}|^2> \propto a(|i-j|)$, can serve as 
good models for 
systems with multifractal eigenfunctions \cite{mfdqs,al,izra}; 
here the function $a(r)$ 
decays with $r$ either as a power law or exponentially (or faster) 
at $r>>1$.
However various other RBM type structures appear in many other physical 
contexts, for example, nuclei, atoms \cite{fggk}, 
solid state \cite{fcic}, quantum chaos \cite{pr}. 
One such example is that of WRBM (W stands for Wigner) \cite{wign} where 
the mean value 
of diagonal elements 
increases linearly along the main diagonal ($<H_{nn}>=\alpha n$). These 
matrices have been shown to be a good model, for example,
 for the tight-binding 
Hamiltonian of a quantum particle in a 1-D disordered system subject to 
a constant electric field as well as for heavy atoms and nuclei 
\cite{fcic}. Another 
example is of the ensemble of banded matrices with diagonal elements 
fluctuating much stronger than the off-diagonal ones 
($<|H_{ii}|^2>/<|H_{ij}|^2> \propto b >>1$ ), also known as RBM with 
preferential basis.  A very interesting problem 
of two interacting particles propagating in a quenched random potential 
can effectively be mapped on to this class \cite {shep}. 
 
 The nature of a localized dynamics leaves its imprints on the 
distribution properties 
of the matrix elements of the generator of motion e.g. Hamiltonian.  
It is therefore desirable to study RBME with various distributions  
and recently a lot of effort has been applied in this direction.  
However the techniques suggested so far \cite{mfdqs,fcic,fmg} depend 
strongly on 
the type of distribution  
chosen and mostly give approximate results. 
 Our aim in this paper is to suggest  a 
method which leads to results in a generic form vaild for most of the  
RBM models. This we achieve by  mapping  
 the problem to the study of a particular class of transition ensembles,  
the one arising during the Poisson $\rightarrow$ standard Gaussian  
transitions  under a perturbation taken from the standard 
Gaussian ensembles (SGE). The  pre-existing 
information regarding the correlations in the latter case therefore can help 
us to obtain the same for RBME.

We proceed as follows. 
Our interest is in the evolution of the eigenvalue distribution of 
$H$, taken from an ensemble of hermitian matrices, due to variation 
of the variances of its matrix elements. We choose the distribution 
$\rho (H)$ of matrix $H$ to be a Gaussian,  
 $\rho (H,y)=C{\rm exp}({-\sum_{k\leq l}
\alpha_{kl} H_{kl}^2 })$ with $C$ as the normalization constant  
and $y$ as the matrix 
 of relative variances 
$y_{kl}={\alpha_{kl}g_{kl}\over \alpha_{kk}g_{kk}}$.   
Let $P(\{\mu_i\},y|H_0)$ be the probability of finding eigenvalues 
$\lambda_i $  of $H$ between $\mu_i$ and $\mu_i+{\rm d}\mu_i$ at 
a given $y$ (with $H_0$ as an initial condition),
\begin{eqnarray}
P(\{\mu_i\},y|H_0)= \int
\prod_{i=1}^{N}\delta(\mu_i-\lambda_i) \rho (H,y){\rm d}H 
\end{eqnarray}
 As the $\alpha$-dependence of $P$ 
in eq.(1) enters only through $\rho(H)$
( ${\partial \rho (H)\over\partial \alpha_{kl}} 
=\left[ (2 \alpha_{kl})^{-1} -H_{kl}^2\right]\rho(H)$),  this equality  
followed by a repeated use of the partial integration alongwith  
eigenvalue equation $H=O^T\Lambda O$, with $\Lambda$ as the eigenvalue 
matrix and $O$ an orthogonal matrix, leads to following, 

\begin{eqnarray}
2{\partial P(\{\mu_i\},y)\over\partial \alpha_{kl}} = 
{1\over \alpha^2_{kl}} I_{kl} 
\end{eqnarray}

where 

\begin{eqnarray}
I_{kl}=\sum_n {\partial \over \partial\mu_n}
\int   
{\rm d}H 
\rho(H)
{\partial\over\partial H_{kl}}\left(
\prod_{i}\delta(\mu_i-\lambda_i){O_{nk}O_{nl}}\right)
\end{eqnarray}

Let us first study the case  with same variance for all the 
diagonal matrix elements such that $g_{kl} \alpha_{kl}=\alpha$ 
for $k=l$  ($g_{kl}=1+\delta_{kl}$)   
while keeping the variances (=$\alpha_{kl}^{-1}$)  of the off-diagonals 
 arbitrary. 
By expressing 
$\sum_{k<l}\left({1\over \alpha}-{1\over 
g_{kl}\alpha_{kl}}\right)...={1\over\alpha}\sum_{k\le l}...-
\sum_{k\le l}{1\over g_{kl}\alpha_{kl}}..$ and using eq.(2), we 
obtain the following relation

\begin{eqnarray}
2\sum_{k<l}\left(1-y_{kl}\right) y_{kl}
{\partial P\over\partial y_{kl}}  
=\sum_{k\le l} {1\over g_{kl}\alpha_{kl}} I_{kl}-
{1\over \alpha}\sum_{k\le l} I_{kl}
\end{eqnarray}

The left hand side of above equation, summing only over all distinct 
$\alpha_{kl}$, can be rewritten as 
${\partial P\over\partial Y}$ with $Y$ given by the condition that 
${\partial \over\partial Y} 
=2\sum_{k<l} y_{kl}(1-y_{kl}){\partial \over\partial y_{kl}}$.   
The first term on the right hand side (eq.(4)) can further be simplified by  
first using a partial integration and subsequently   
the eigenvalue equation for $H$: 
 $\sum_{k\le l} {1\over \alpha_{kl}} I_{kl}=
\sum_{n}{\partial \over \partial\mu_n}\left(\mu_n P\right)$.
The second term can similarly be 
reduced by differentiating the terms inside the brackets 
in (3) 
giving us two integrals. The use of orthogonality relation 
of matrix $O$ in the first integral so obtained and 
 the equality 
$\sum_{k\le l} {\partial O_{nk} O_{nl}\over \partial H_{kl}}=\sum_m 
{1\over \lambda_m-\lambda_n}$ in the second \cite{ps} leads to a 
F-P equation 

\begin{eqnarray}
{\partial P\over\partial Y} &=& 
\sum_{n}{\partial \over \partial\mu_n}\left(\mu_n P\right) \nonumber \\
&+&{1\over \alpha}\sum_n {\partial \over \partial\mu_n}
\left[ {\partial \over \partial\mu_n} +
\sum_{m\not=n}{\beta \over {\mu_n-\mu_m}}\right] P 
\end{eqnarray}

where $\beta=1$.  
 By using the unitarity of eigenvectors and 
following the same steps, it can be proved for 
complex Hermitian case too (now $\beta=2$).
 It should be noted here, for the later use, that the required form 
of eq.(5) is 
 obtained by applying an appropriate partitioning of sums appearing 
in eq.(4) which leads to rewriting of the terms containing all the unequal 
coefficients $\alpha_{kl}$ as the derivative with respect to parameter $Y$. 
The coefficient appearing with drift and diffusion terms in eq.(5) is the 
one common to many of the matrix elements.

	The definition of $Y$  depends on  the 
relative value of variances of the off-diagonals as 
discussed below for some cases:

{\it Case I.
 When all the off-diagonals have same variances  
such that $\alpha_{kl}=\alpha'$ (for $k\not= l$)} 

In this case, as $y={\alpha'\over \alpha}$, 
therefore $Y= 2\;{\rm log}{y\over |y-1|}$.


{\it Case II.
 When variances of the off-diagonals change with respect to distance 
from the diagonal   
such that $\alpha_{kl}=\alpha_r$ (for $r=|k-l|>b>0$) 
and $\alpha_{kl}=\alpha$ for $r\le b$.}  

Now all the off-diagonals (with $y_r\not=1$)  
contribute separately to $Y$, 
$Y=2 \sum_{r=b+1}^N {\rm log}{y_r\over |y_r-1|}$
where $y_r={\alpha_r \over \alpha}$.

For all $y_r >1$, this represents 
a RBM with strongly fluctuating  diagonal elements  
 (known as RBM with preferential basis). 
	The case with all $y_r<1$ describes the standard RBM, used to model, 
for example, the spectral statistics of the tight binding 
Hamiltonian of a quantum particle in a 1-D system with long range random 
hoppings \cite{mfdqs} and also for quasi 1-D disordered wires \cite{fm91}. 

{\it Case III.
 When the off-diagonals form various groups with different varainces 
(with same variance for each matrix element in one such group).}

Now  contribution to $Y$ comes from each such group. For example, with 
$M$ groups of variances given by $\alpha_g^{-1}$ ($g=1\rightarrow M$, 
$\alpha_g \not=1$),  
$Y= 2 \sum_{g=1}^M {\rm log}{y_{g}\over |y_{g}-1|}$
where $y_g={\alpha_g\over\alpha}$. This will be helpful to 
model the more general cases  where the variance $<|V_{ij}|^2>$ is dependent on both 
 indices $i,j$ instead of their difference $|i-j|$.

	The steady state of eq.(5), 
$P(\{\mu_i\},\infty)\equiv P_{\infty}=\prod_{i<j} |\mu_i-\mu_j|^{\beta}
{\rm e}^{-{\alpha\over 2}\sum_k \mu_k^2}$, 
is  achieved for $Y \rightarrow
\infty$ which corresponds to $y \rightarrow 1$ in case (I),   
 $y_r \rightarrow 1$ in case 
(II)  and $y_{kl} \rightarrow 1$ for the case (III).   
 This indicates that, in the steady 
state limit, system tends to belong to the 
standard Gaussian ensembles.  
	Eq.(5) (later referred as variance-variation or VV case) 
is formally the same as the F-P equation governing the 
Brownian motion of particles in Wigner-Dyon gas \cite{meta} with 
transition parameter 
being the relative variance  
 in the former and time in the latter. This is also similar 
to the F-P equation for the eigenvalue distribution   
of a hermitain matrix $H=H_0+\tau V$ undergoing random perturbation $V$ 
of strength $\tau$ and taken from a SGE with 
arbitrary initial condition $H_0$ (later referred as perturbation variation 
or PV case) \cite{os};  
here $\tau$ acts as the transition parameter.
$P(\{\mu_i\},Y)$ can therefore be obtained by the  
same procedure as used in PV case which is briefly given as follows.  
The transformation 
$\Psi=P/\sqrt{P_\infty}$ 
 reduces eq.(5) to a "Schrodinger equation" form, 
 ${\partial \Psi \over\partial Y} = \hat H \Psi$, 
where the 'Hamiltonian' $\hat H$ turns out to be a Calogeo-Moser 
(CM) Hamiltonian \cite{ap,akhi}, 
$\hat H = 
\sum_i{\partial^2 \over \partial\mu_i^2}
-{1\over 4}\sum_{i<j}{\beta (\beta-2)\over (\mu_i-\mu_j)^2}-
{\alpha^2 \over 4}\sum_i 
\mu_i^2$, 
 and has well-defined eigenstates and eigenvalues \cite{znc}.  
The "state" 
$\Psi$ or $P(\{\mu_i\},Y| H_0)$ can therefore be expressed as a 
sum over its 
eigenvalues and eigenfunctions  which on integration over all the 
initial conditions  $H_0$ 
leads to the joint probability distribution $P(\{\mu_i\},Y)$ and 
thereby correlations. Unfortunately, due to technical problems, 
 the latter could be evaluated only for the transitions with final 
steady state (limit $\Lambda\rightarrow \infty$) as GUE \cite{akhi}.  
However a set of hierarchic relations among the correlators for all 
transitions in PV case, and therefore for VV case, can be  obtained 
by a direct integration of the F-P equation for $P$ (eq.(5)) \cite{ap,akhi}. 
For example, in large $N$-limit, the evolution of level density $\rho(\mu,Y)$ 
is governed by the Dyson-Pastur equation \cite{ap,akhi} which results in  
a semi-circular form for $\rho$ (thus agreeing well with the $\rho$ obtained in 
\cite{klh} by super-symmetry technique).  The relation 
$\rho(\mu,Y)=N^{-1}\sum_{n} \rho(\mu,n,Y)$ can further be used to 
obtain the equation for the local density of states 
(LDOS) $\rho(\mu,n,Y) $, 
a more informative and experimentally accessible quantity (referred 
below as $\rho_n$), 

\begin{eqnarray}
{\partial \rho_n\over\partial Y} = 
-\beta {\partial \over \partial \mu}\left( 
\sum_m {\bf P}\int {\rm d}\mu'
{\rho_m \over {\mu-\mu'}}\right) \rho_n. 
\end{eqnarray}
This nonlinear equation gives uniquely $\rho(E,n,Y)$ starting from 
an initial $\rho(E,n,0)$.

	For localization studies, it is appropriate to choose the initial 
 ensemble $H_0$ as that of diagonal matrices ($P(H_0)
 \propto {\rm e}^{-{\alpha\over 2}\sum_i H_{ii}^2}$ with $V_{ii}=\mu_{0i}$ and   
all $y\rightarrow \infty$) which corresponds to Poisson distribution for 
eigenvalues and $Y=0$.
 With  equilibrium  
distribution ($Y\rightarrow \infty$) given by SGE,  
this case thus 
represents  a Poisson $\rightarrow$ SG transition  
 with $Y$  
  as a transition parameter  and  the 
intermediate ensembles representing various RBMEs 
depending on the type of $y_r$'s. Thus the correlations here will be  
  similar to those in the Poisson$\rightarrow$SG transition in SGE  
(PV case with  $H_0$ taken from a Poisson  ensemble). 
 We already know that the transition for PV case is abrupt 
 for large dimensions and finite $\tau$ and a rescaling of 
$\tau$ by mean spacing $D^2 (\propto N)$ is required 
to make it smooth, the new transition parameter being 
$\Lambda={\tau  \over D^2}$.  A similar rescaling  
 should also be applied to $Y$ in VV case for the same reason; here too 
 $D^2 \propto N$ \cite{klh}. (This $N$-dependence of $D$ also  
follows from analogy of the evolution-equations for level-densities in the 
two cases ).  

 Fortunately the two-point correlation $R_2(r;\Lambda)$ for 
Poisson $\rightarrow$ GUE transition in PV case 
 has already been obtained \cite{akhi} and, as discussed above, 
is also valid for the VV case (now $\Lambda=Y/D^2$)   
.

\begin{eqnarray}
 R_2 (r;\Lambda) - R_2(r;\infty)={4\over \pi}\int_0^\infty {\rm d}x 
\int_{-1}^1 {\rm d}z \;{\rm cos}(2\pi rx) 
\;{\rm exp}\left[-8\pi^2\Lambda x(1+x+2z\sqrt x)\right]
\left({\sqrt{(1-z^2)}(1+2z \sqrt x) \over 1+x+2z \sqrt x}\right)
\end{eqnarray}

where $R_2(r,\infty)=1-{{\rm sin}^2(\pi r)\over \pi^2 r^2}$ (the GUE limit). 
As can easily be checked,  
above equation has the correct limiting behaviour, 
that is, $R_2= 0$ for $\Lambda \rightarrow 0$ (the Poisson case $Y=0$) and 
$R_2=R_2(r;\infty)$ for $\Lambda \rightarrow \infty$.  As obvious from 
eq.(7), $R_2$ 
for intermediate ensembles will depend on definition of $Y$ and 
therefore on the nature of localization which results in various types of 
level-statistics. 
For example, let us calculate 
the correlation for one such case, namely, $H_{ij}=G_{ij} a(|i-j|)$ with 
$G$ a typical member of SGE and $a(r)=1$ and $(b/r)^\sigma$ for 
$r\le b$ and $> b$ ($b>>1)$ respectively. 
For this case, $y_r=({r\over b })^\sigma$ and 
$Y \approx 2 \sum_{r=b+1}^N \left({b\over r}\right)^{2\sigma}$ 
 which gives  
$\Lambda \propto N^{2-2\sigma}$.
 For $\sigma >1$ with $N\rightarrow \infty$, therefore, the 
eigenvalue statistics approches Poisson limit, $\Lambda$ being very small. 
Similarly for $\sigma <1$, $\Lambda$ is sufficiently large and the 
eigenvalue statistics approches SG limit. For $\sigma=1$, the 
independence of $\Lambda$ from $N$ leads to an eigenvalue statistics, 
very different 
from that of SGE or Poisson and therefore agrees well with the results 
obtained in \cite{mfdqs} by using non-linear $\sigma$-model technique. Further 
 $R_2$, (eq.(7)) is also in good accordance 
with the one given in \cite{mfdqs} (which can be seen by a 
direct substitution of eq.(52) of \cite{mfdqs} in eq.(17) 
of \cite{akhi}).  

	It is possible to have physical situations when the matrix 
elements form various groups such that those in one group have the same 
value for $\alpha_{kl}g_{kl}=\alpha_r$ ($r=1\rightarrow M$ with $M$ as 
total number of groups). This case can similarly be treated by using  	
2$\sum_{r=1}^M \left({1\over \alpha_s}-{1\over 
\alpha_r}\right)\alpha^2_r
{\partial P\over \partial \alpha_r}
={1\over\alpha_s}\sum_{r} I_r-
\sum_r {1\over \alpha_r} I_r $ with 
$I_r=\sum_{\{k,l\}\epsilon r} I_{kl}$ and $I_{kl}$ still given 
by eq.(3). Here $\alpha_s$ refers to the $\alpha$-coefficient of any 
one (chosen arbitrarily) of the $M$ groups. 
This partitioning again leads to eq.(5) (with 
$\alpha \rightarrow \alpha_s$ on the right hand side) and therefore 
similar level-correlations in terms of the transition parameter 
$Y= 2\;{\rm log}\prod_r {y_r\over |y_r-1|}$ with 
$y_r={\alpha_r\over \alpha_s}$.  

	Finally let us consider the case of WRBM  when mean 
value of the diagonal elements increases  along the main diagonal: 
$<H_{nn}>=\gamma f(n)$. For simplification, we still take $\rho(H)$ 
to be a Gaussian: 
$\rho(H)=C {\rm e}^{-\sum_{k\le l}
\alpha_{kl}(H_{kl}-\gamma f(k)\delta_{kl})^2}$ with all distinct 
$\alpha_{kl}$.  
Proceeding as before, one again obtains eq.(2)  with $I_{kl}$ 
given by eq.(3) but now, instead of applying the partitioning of the 
sum, we just evaluate  
$2 \sum_{k\le l}{\partial P\over \partial \alpha_{kl}}\alpha^2_{kl}g_{kl}$.
This leads  to a F-P equation without the linear drift term, 
\begin{eqnarray}
{\partial P(\{\mu_i\},Y)\over\partial Y} &=& 
\sum_n {\partial \over \partial\mu_n}
\left[ {\partial \over \partial\mu_n} +
\sum_{m\not=n}{\beta \over {\mu_m-\mu_n}}\right] P 
\end{eqnarray}
where $Y=\sum_{k\le l} (2 \alpha_{kl} g_{kl})^{-1}$ and the steady state 
$P_{\infty}=\prod_{i<j}|\mu_i-\mu_j|^{\beta}$ is obtained in limit 
$Y\rightarrow \infty$ or $\alpha_{kl}\rightarrow 0$. 
Similarly for the degenerate case, 
when all the $\alpha_{kl}$ can be divided into $M$ groups with distinct 
values $\alpha_r=\alpha_{kl} g_{kl}$, $r=1\rightarrow M$, 
$2 \sum_{r=1}^M {\partial P\over \partial \alpha_r} \alpha_r^2$  can be used 
to obtain eq.(8) where now $Y=\sum_r (2 \alpha_r^{-1})$.  
 An application of the same transformation 
$\Psi=P/\sqrt{P_{\infty}}$ reduces eq.(8) in the "Schroedinger 
equation" form with Hamiltonian $\hat H$, 
$\hat H = 
\sum_i{\partial^2 \over \partial\mu_i^2}
-{1\over 4}\sum_{i<j}{\beta (\beta-2)\over (\mu_i-\mu_j)^2}$, 
which is same as CM Hamiltonian except for the absence of confining 
potential and, as shown in \cite{akhi}, has essentially the same solution for 
$\Psi$ (without the Gaussian factor) and therefore $P(\mu,Y|H_0)$. 
However, for the evaluation 
of $P(\mu,Y)$, one needs to integrate over all the initial conditions 
through which the non-zero mean values of the diagonals enter in the 
calculation and may influence the statistics for large $\gamma$-values. 
Nontheless, in large $N$-limit, the hierarchic relations among the 
correlators are of the 
 same type as in other RBM cases \cite{ap,akhi} with  LDOS 
 satisfying  the eq.(6) \cite{akhi}. 
Here the choice of $H_0$    
from a Poisson ensemble 
($\rho(H_0)\propto {\rm e}^{-{\alpha\over 2}\sum_j (H_{jj}-\gamma f(j))^2}$) 
gives the initial value of $Y=\alpha^{-1}$. 
With $Y\rightarrow\infty$ corresponding to a SG type ensemble, therefore, 
 WRBM also appear as the intermediate ensembles in the  
Poisson $\rightarrow$ SG type transition.

	In this paper, we have analytically studied the response of energy 
levels of complex quantum systems to the changing distribution of matrix 
elements of their hamiltonians.  
 Our results indicate that the $n^{\rm th}$ order correlations at a given 
variance are the same as the corresponding eigenvalue-correlations of  
hamiltonian $H=H_0+\tau V$ at a given $\tau$ value with $H_0$ taken from 
a Poisson ensemble and $V$ taken from a SG ensembles. This analogy also 
extends to the $2^{\rm nd}$ order parametric correlators 
 however the Markovian nature of F-P equation restricts 
 from making similar conclusions about higher orders.  
The intermediate ensembles arising in the VV transition 
correspond to various types of RMEs, and 
 as our mapping suggests, the results for most of them 
 can be obtained by studying just one transition in full detail, 
namely, an appropriate initial ensemble $\rightarrow$ SGE caused due to a SG perturbation;
 the knowledge of their correlations can help 
 us in statistical studies of many important physical properties of 
 complex systems.  
Further, as discussed in \cite{km}, 
some of the RBMEs can also be connected to various other types of ensembles; 
a knowledge of the properties of former will therefore help in 
statistical analysis of the latter. It should also be possible to apply  
this technique to the ensembles of Unitary matrices (periodic RBMs) as well. 

	Our study still leaves many questions unanswered. At present, 
it is not clear whether our method can  be extended to non-Gaussian 
distributions as well, at least in some limit. We are also unable to say whether  
a similar mapping can also be done for the eigenvector statistics. 
In this connection, however, it should be mentioned that analogy of equations 
governing the eigenvalue distribution leads to a similar form of equations 
for the distribution of matrix elements \cite{meta} and, 
therefore, it should somehow manifest itself in eigenvector distributions too.

I am greatful to B.S.Shastry, B.L.Altshuler and V.Kravtsov    
for useful suggestions during the course of this study.

\end{document}